\documentclass[twocolumn]{mnras}
\usepackage{amsmath,amstext}
\usepackage{graphicx}

\def\modified{\empty}
\def\teff{$T_{\rm eff}$}

\title[The sub-Jovian desert of exoplanets]{The sub-Jovian desert of exoplanets: parameter dependent boundaries and implications on planet formation}

\author[Gy. M. Szab\'o and Sz. K\'alm\'an]{
Gy.~M. Szab\'o{$^1,^2$}\thanks{E-mail: szgy@gothard.hu}
Sz. K\'alm\'an{$^3$}
\\
{$^1$}{ELTE E\"otv\"os Lor\'and University, Gothard Astrophysical Observatory, 9700 Szombathely, Szent Imre h. u. 112, Hungary}\\
{$^2$}{MTA-ELTE Exoplanet Research Group, 9700 Szombathely, Szent Imre h. u. 112, Hungary}\\
{$^3$}{University of Szeged, Institute of Physics, D\'om t\'er 9, Szeged 6720, Hungary}
}

\date{Accepted XXX. Received YYY; in original form ZZZ}
\pubyear{2019}
\begin{document}
\label{firstpage}
\pagerange{\pageref{firstpage}--\pageref{lastpage}}
\maketitle


\begin{abstract}
The period-mass and period-radius distribution of exoplanets is known to exhibit a desert. Unlike the existence of very hot ($P_{\rm orb}<3$~d) super-Earths and hot Jupiters, no planets are known between a super Earths and sub-Jupiters with as short orbital periods as a day or two.
In this Letter, we show that the period boundary of this desert is dependent on stellar parameters ($T_{\rm eff}$, $[M/H]$, $\log g$ in the order of significance), there is a conclusive dependence on the incident stellar irradiation, and a dependence on the stellar mass acting only on planets around $T_{\rm eff}<5600$~K host stars. We found a significant lack of very inflated planets on closest orbits to the host star. There is no significant dependence on tidal forces currently acting on the planet, planet's surface gravity, or current filling factor of Roche lobe. These distributions are most compatible with the dominant role of photoevaporation in forming the desert.
\end{abstract}


\section{Introduction}
We had reported a lack of intermediate mass planets ($0.02 M_J < M_p < 0.8 M_J$) at short periods (approximately $P< 3$~d) in Szab\'o and Kiss (2011), which was initially called as the ``sub-Jupiter desert'', since it censored out mostly sub-Jupiters and Neptunes. The term ``Neptune desert'' was also used (Mazeh et al. 2016), who confirmed the presence of this desert, and also gave analytic formulae of its boundaries. 

In the past years, it has been argued whether the desert is completely empty or has a low population of planets. The desert was known to host a low number of planet candidates, but if the sample is restricted to dynamically confirmed planets only, the desert remained completely empty (Morton et al. 2016, Owen and Lai, 2018). However, {\it Kepler} planets in the desert have been identified by the California-{\it Kepler} survey (Fulton \&{} Petigura, 2018), and NGTS-4b was also found to be in the desert (West et al. 2018). By all means, the evidence suggests that the presence of planets in the desert is very much reduced by some scenario at some point of the evolution. Either the formation of planets in the desert is prohibited, or they start forming, but rapidly migrate outwards or into the star to clear out the desert, or the planets in the desert are rapidly built up to form a Hot Jupiter, or ruined down by evaporation down to a super Earth. 

Several possible scenarios have been invoked to explain the origin of this desert, see e.g Owen and Lai (2018) for a comparative summary.  A group of explanations point toward the effective acting of photoevaporation on sub-Jupiters. Lopez and Fortney (2013) predicted an underpopulated sub-Neptune region due to varying internal composition of the planets of different classes. This points towards the parameter-dependent position of the boundaries of the desert.  Indeed, an overpopulation of planets around higher metallicity stars has been found at the lower boundary of the desert  (Dong et al. 2018; Petigura et al. 2018), which can imply a metallicity-dependent photoevaporation (Owen \&{} Murray-Clay 2018). Based on the positions of WASP-151b, WASP-153b, WASP-156b at the upper boundaries of the desert, Demangeon et al. (2018) suggested that the ultra-violet irradiation plays an important role in this depletion of planets observed in the exoplanet population. However, Ionov et al. (2018) concluded that the desert of short-period Neptunes could not be entirely explained by evaporation of planet atmosphere caused by the radiation from a host star. 

Also, migration scenarios have been considered to explain the desert. In the high-eccentricity migration scenarios, the hot sub-Jupiter orbits outside the desert initially, when its eccentricity is suddenly excited by a scattering event, such as an encounter of another planet, or a stellar encounter to the system. Then the orbit undergoes a circularisation, and will be relaxed closer to the boundary the the desert. 

Owen \&{} Lai (2018) gives a detailed explanation for the position of the desert, accounting for tidal boundaries, high-eccentricity evolution with tidal decay, and stellar irradiation. They succeeded in reproducing the position and the shape of the desert, and predicted that the boundaries of the desert depend on the core mass of the planet: planets with larger core masses can go deeper into the desert at the period boundary. They predicted a metallicity-dependent border: planets with more metallicity can appear at shorter periods. 

The third family of explanations are based on processes in the evaporating inner disk boundary. Armitage (2007) predicted that due to tidal interactions between the inner edge of the gap and the planet, the planet will be trapped for a while, and will be driven out as the inner edge is moving outward. In a later work, Alexander \&{} Armitage (2009) showed that in this process, hot Neptunes can evolve into hot Jupiters, which later can evolve to closer orbits to the star, leading to the clump of very hot Jupiters at the upper boundary of the desert. The predicted morphology is very similar to what we see in the distribution of exoplanets.

In this paper, we demonstrate that the boundary of the desert significantly depends on a few fundamental parameters of the host star and the planet. 

\section{Methods}

We explored the planet occurrence near the boundaries of the desert, as a function of various system parameters. We used the database of confirmed exoplanets.\footnote{https://exoplanetarchive.ipac.caltech.edu, downloaded on 20 July 2018.} We filtered the table for those planets where the period, planet radius, planet mass, and the effective temperature of the host star were known, resulting in 607 planets. In this sample, 550 planets had [M/H] metallicity measurement, and 406 were completed with stellar mass $M_*$ and semi-major axis $a$, which are factors in the dynamical parameters. For each investigations, we used the largest complete subset of confirmed exoplanets.

{We were looking for boundary effects in a test region which covers well the desert and its boundary.} The desert reaches the deepest extension into the period regime  in the size range of 0.28 to 0.63 Jupiter radii (Fig. 1). {By simply following the visual impression, we defined} our main test region in the size range of 0.28--0.63~$R_J$, marked by $A$ in Fig. 1. We also defined a control region just below the the desert, the sample of Earths between 0.06--0.16 $R_J$ (marked by $B$ in Fig. 1).

\begin{figure*}
\vskip0mm%
\hskip17mm\includegraphics[viewport=143 261 511 618, width=4.3cm]{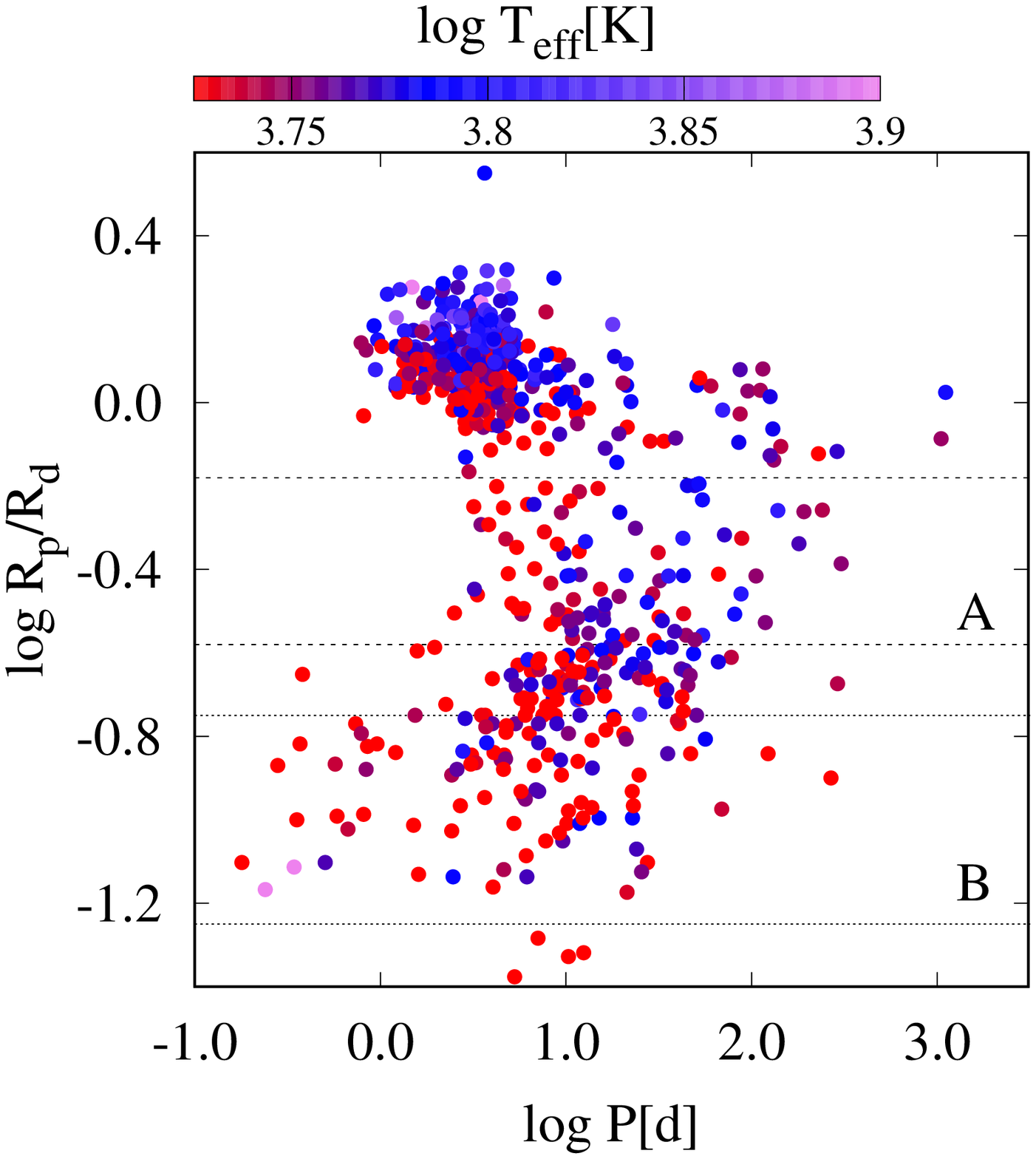}\hskip15mm%
\includegraphics[viewport=143 261 511 618, width=4.3cm]{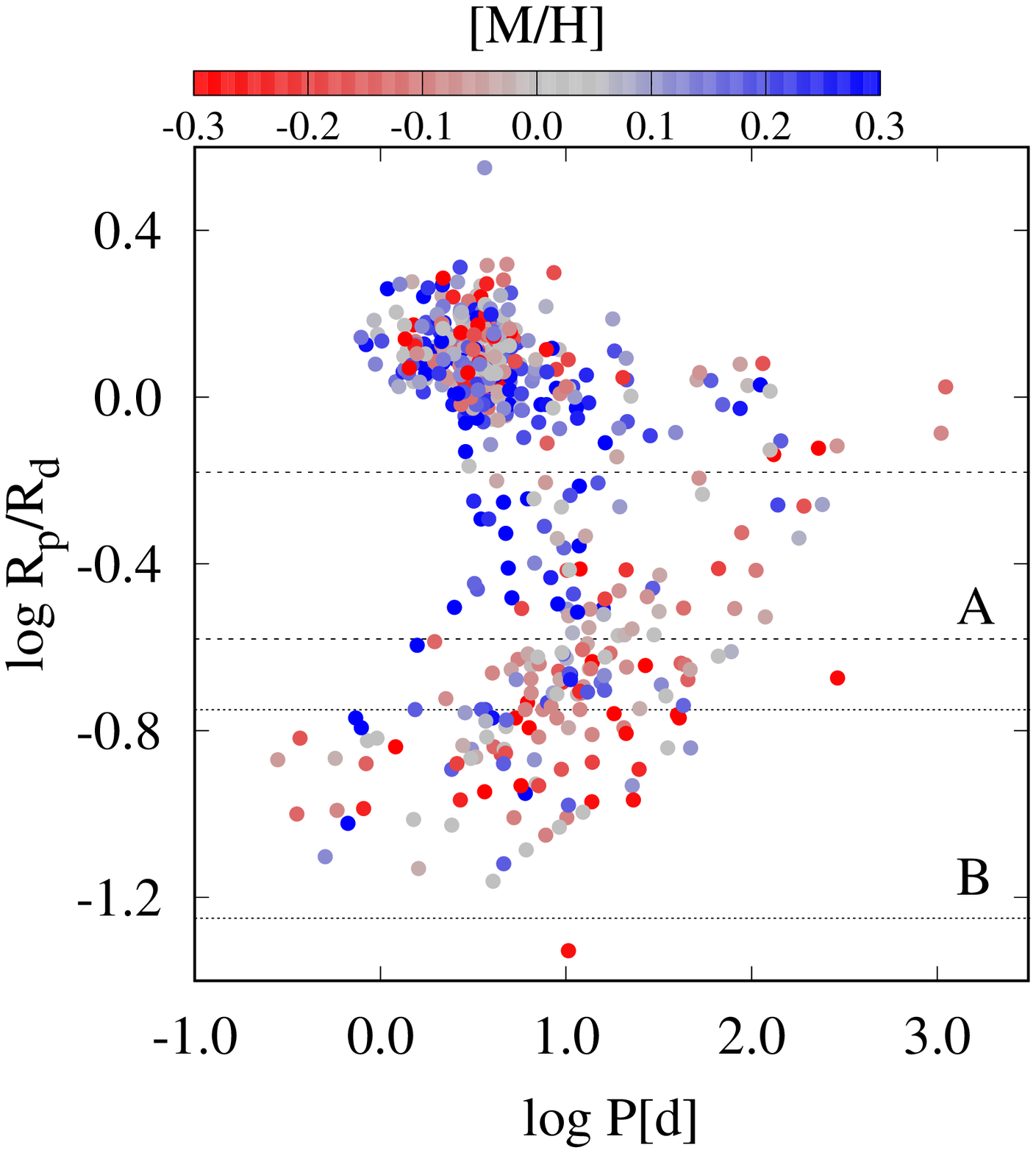}\hskip15mm%
\includegraphics[viewport=143 261 511 618,  width=4.3cm]{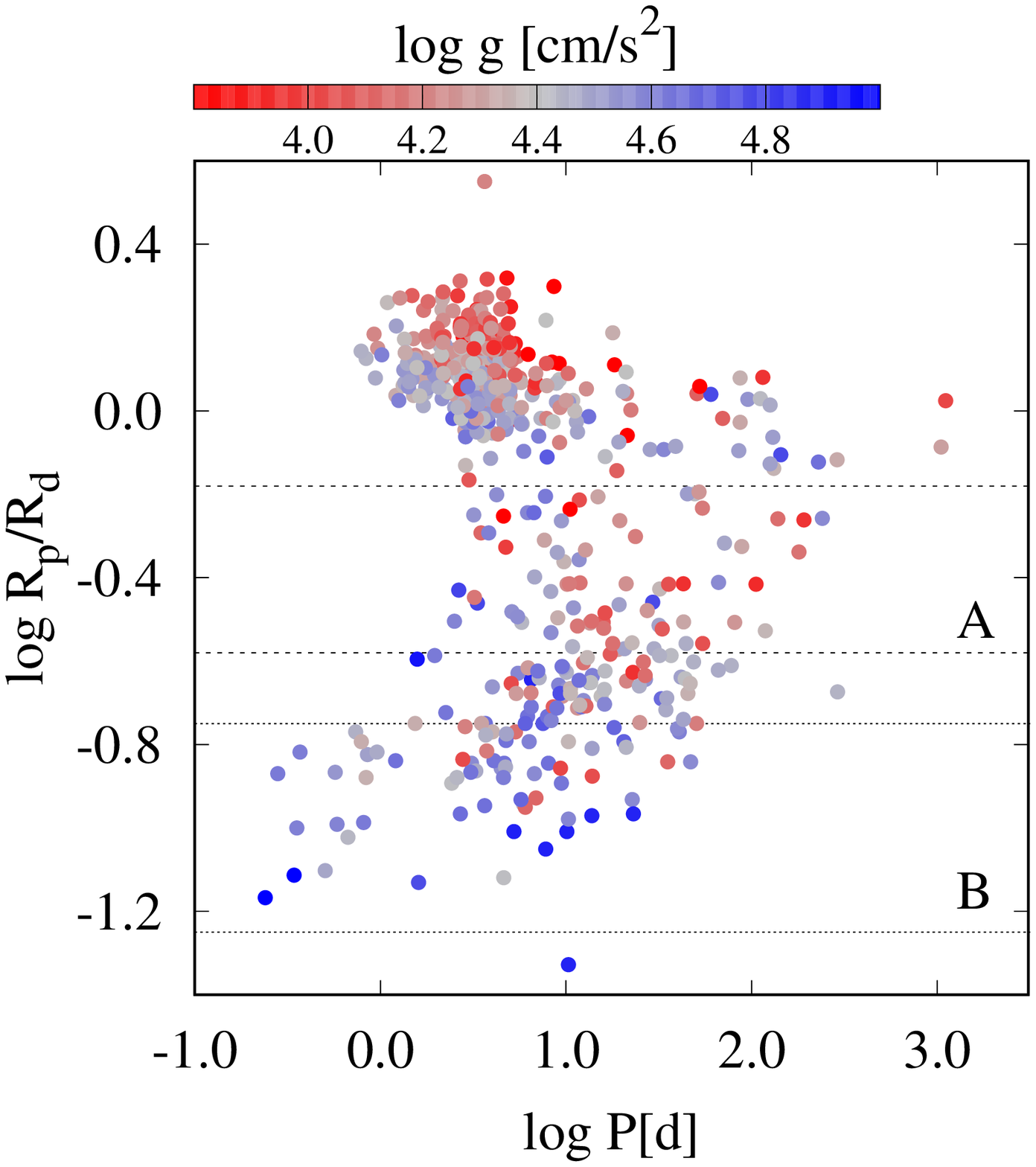}%
\vskip18mm%
\hskip6.0mm\includegraphics[viewport=107 215 396 323,width=4.3cm]{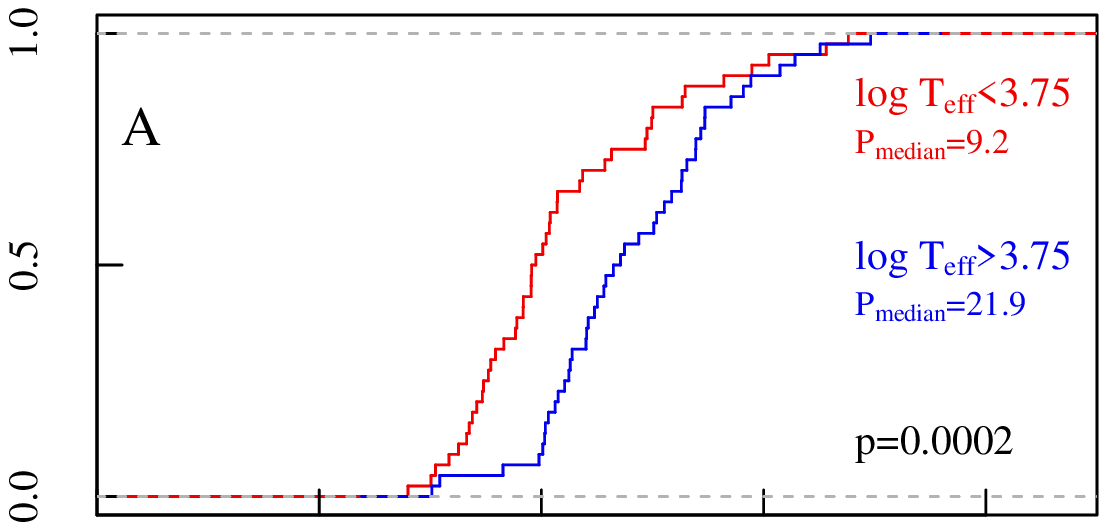}\hskip15mm%
\includegraphics[viewport=107 215 396 323,width=4.3cm]{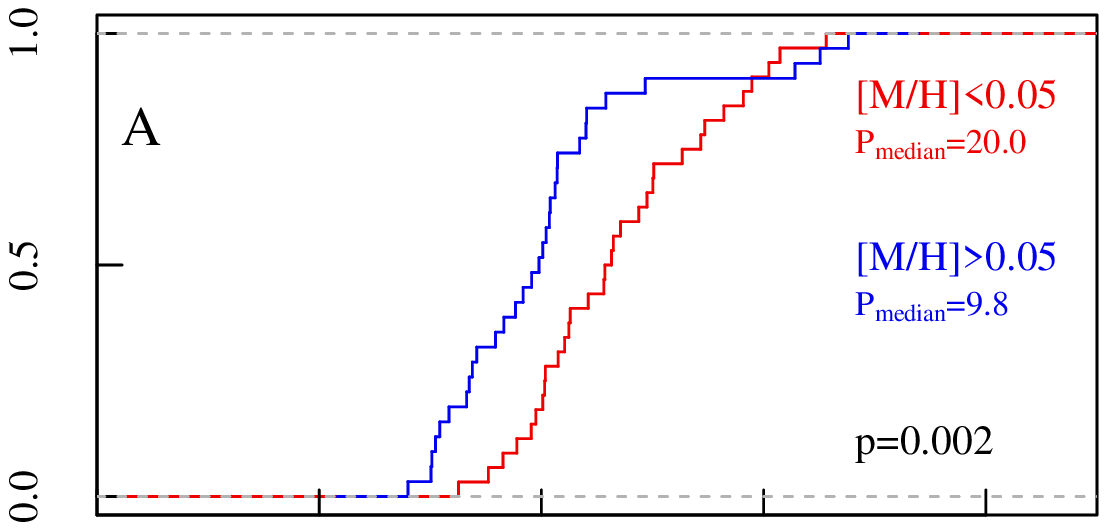}\hskip15mm%
\includegraphics[viewport=107 215 396 323,width=4.3cm]{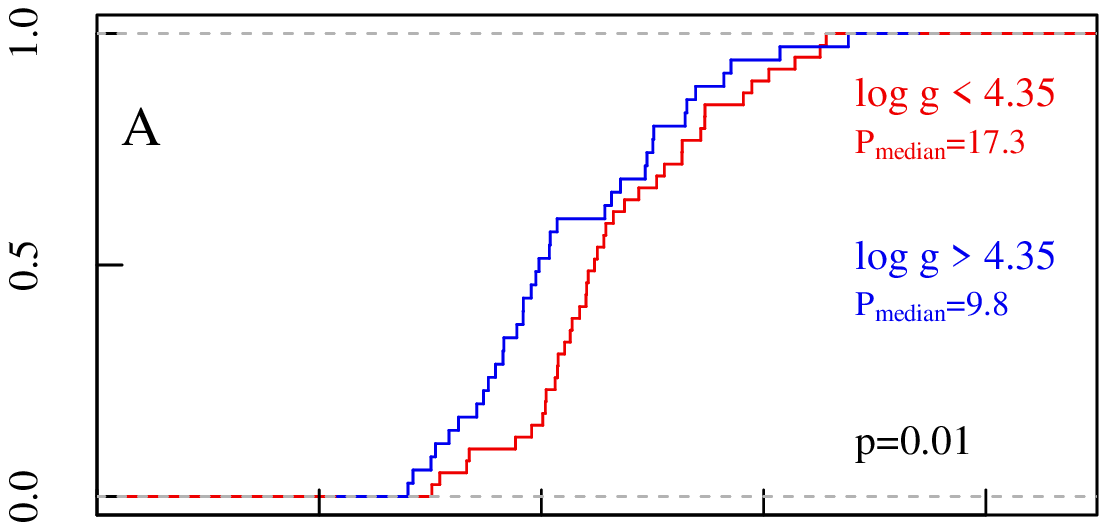}\hskip0mm%
\vskip8mm
\hskip6.0mm\includegraphics[viewport=107 181 396 323,width=4.3cm]{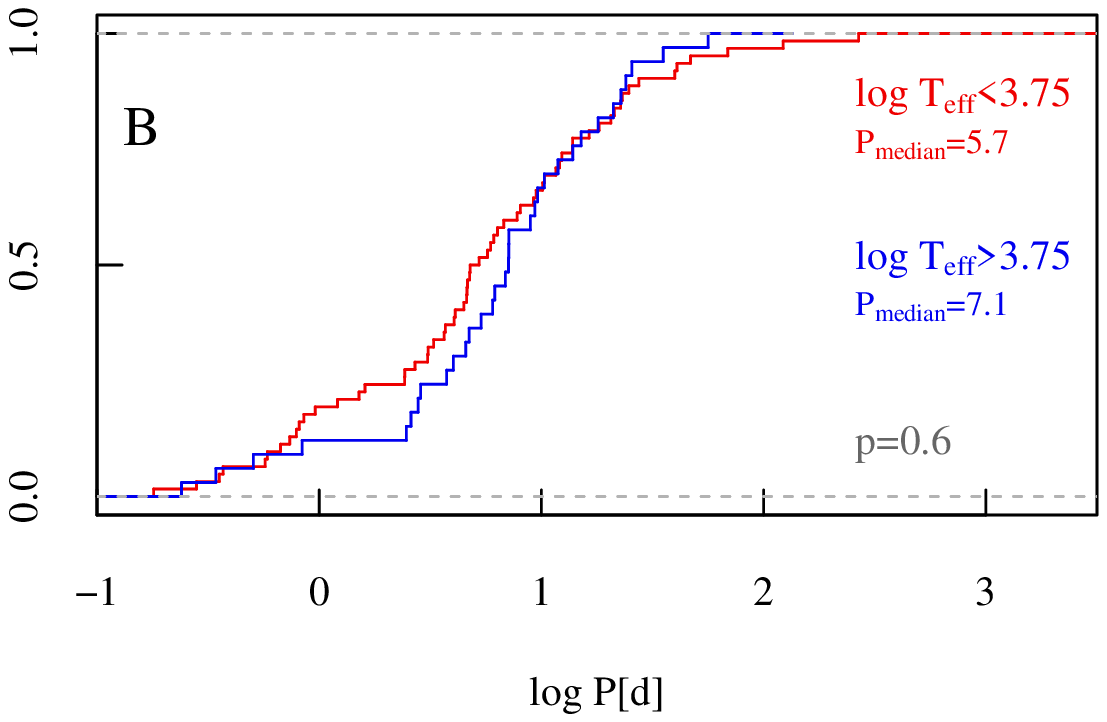}\hskip15mm%
\includegraphics[viewport=107 181 396 323,width=4.3cm]{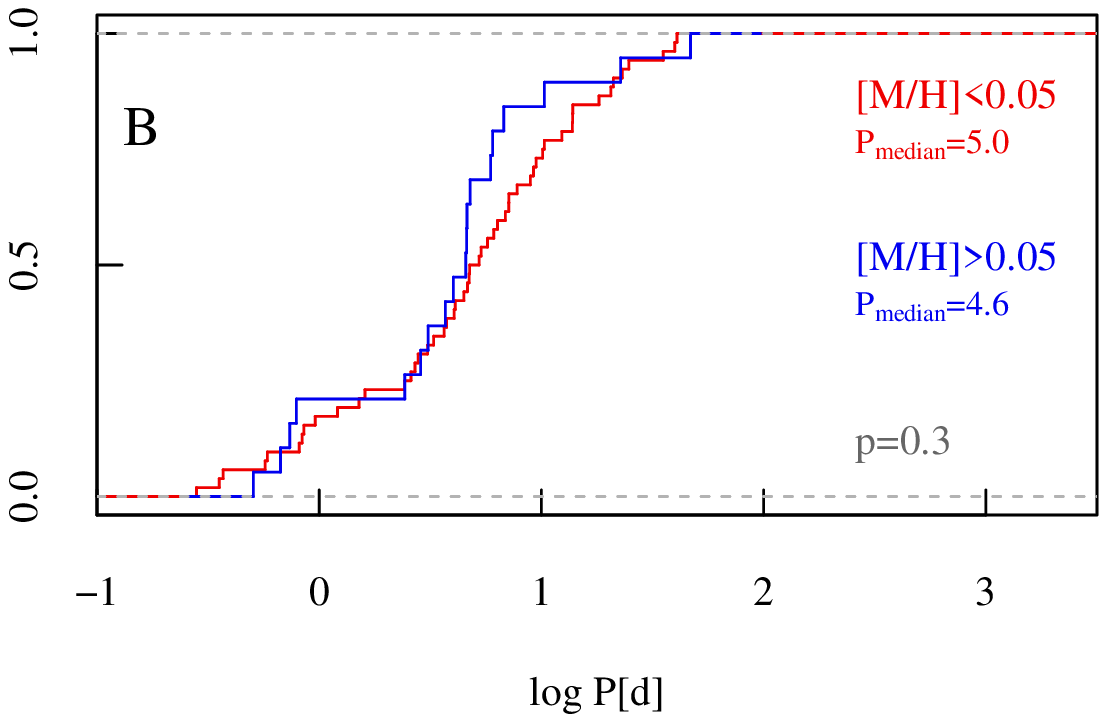}\hskip15mm%
\includegraphics[viewport=107 181 396 323,width=4.3cm]{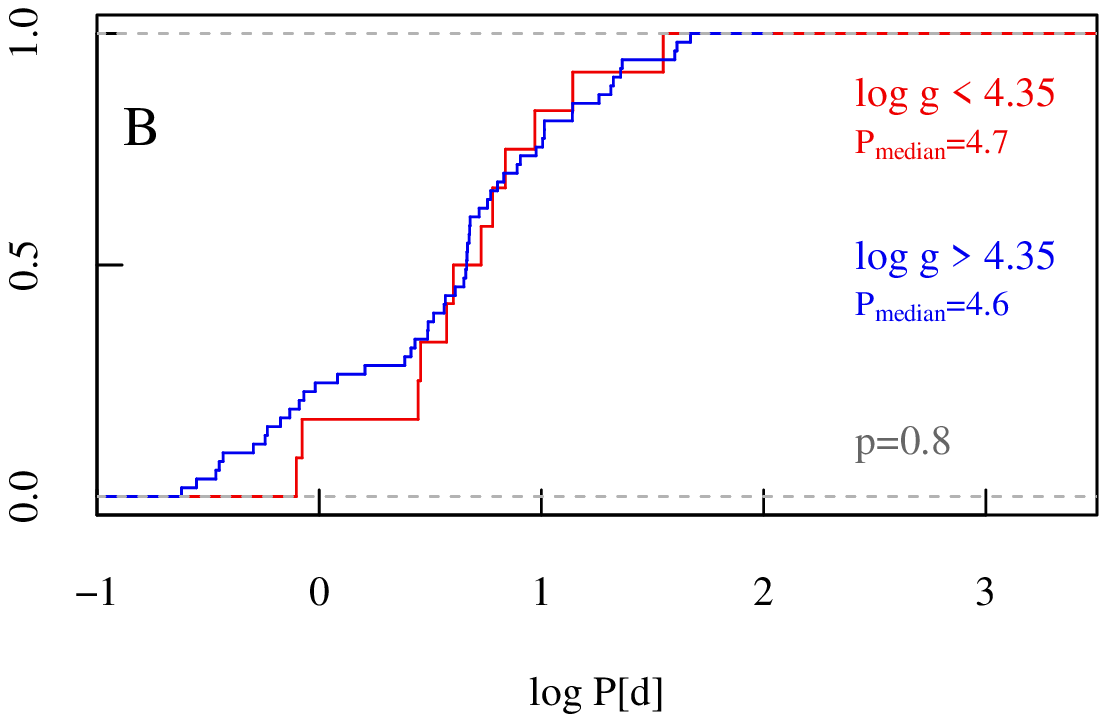}%
\vskip10mm
\caption{Significant parameter dependences of the boundaries of the sub-Jupiter desert: stellar \teff{} temperature (left column), $[M/H]$ metallicity (middle column) and $\log g$ (right column). Top rows: the exoplanets in the period--radius plane, colored by the parameter values. The two regions denoted by A and B are also shown. Middle rows: The cumulative distributions of the orbital period in the A region, comparing subsamples distinguished by high and low values of the examined parameter. The $p$ value of the K-S statistics is also given in the plot. Bottom rows: The same as the middle row, but for the B region.}
\end{figure*}

At the boundaries of the desert, we have have examined many scatterplots showing the period-radius plane, with symbol colors coding the third parameter which we tested to influence the planet occurrence at the boundary of the desert. We involved the stellar effective temperature $T_eff$, stellar mass $M_*$, stellar $\log g$, stellar mass $M_*$, the stellar metallicity $[M/H]$, the equilibrium temperature of the planet $T_B$, the mean density of the planet relative to that of Jupiter, $\rho_p/\rho_J$, the filling factor of the Roche lobe, and the factor scaling current tidal forces on the planet $M_*/a_p^3$. The shape of the distributions tell us how an increased stellar temperature affects the extension of the desert. 

{We had split the sample at roughly the median of the examined parameters, and compared the period distribution of planets with low/high values. To be quantitative, we made use of a Kolmogorov-Smirnov test (KS-test, see e.g. Feigelson and Babu, 2011), which tells us whether two samples are derived from different distributions. We selected a system parameter, e.g. temperature, and if we found that the period distribution of planets in the $A$ sample was different around hot and colder stars, then this was an evidence for the temperature-dependence. If the detected dependence was not more present in the $B$ region, it is an evidence for its specificity at the desert boundary.
}

\section{Results}

\begin{figure*}
\hskip17mm\includegraphics[viewport=143 261 511 618, width=4.3cm]{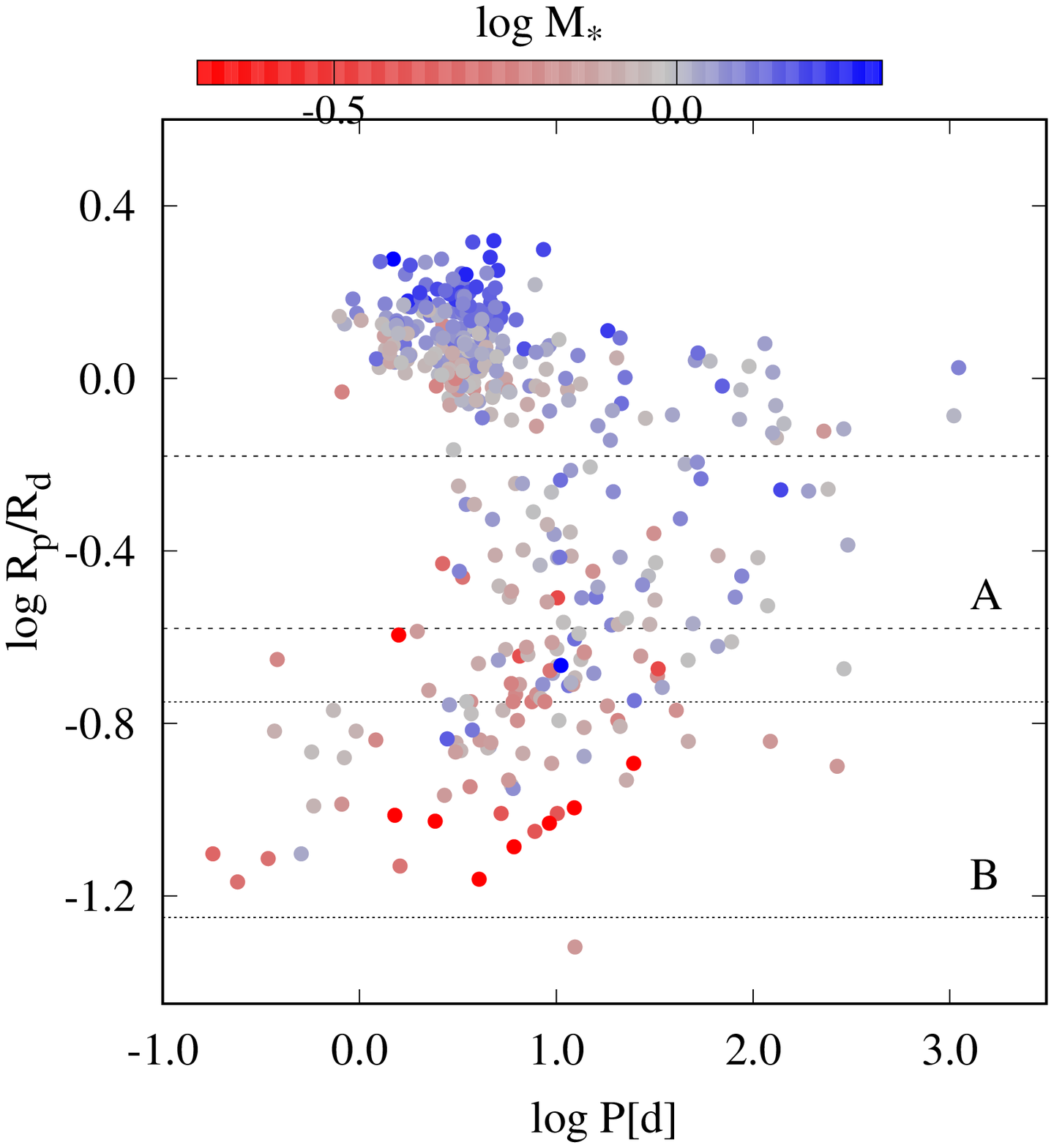}\hskip15mm%
\includegraphics[viewport=143 261 511 618, width=4.3cm]{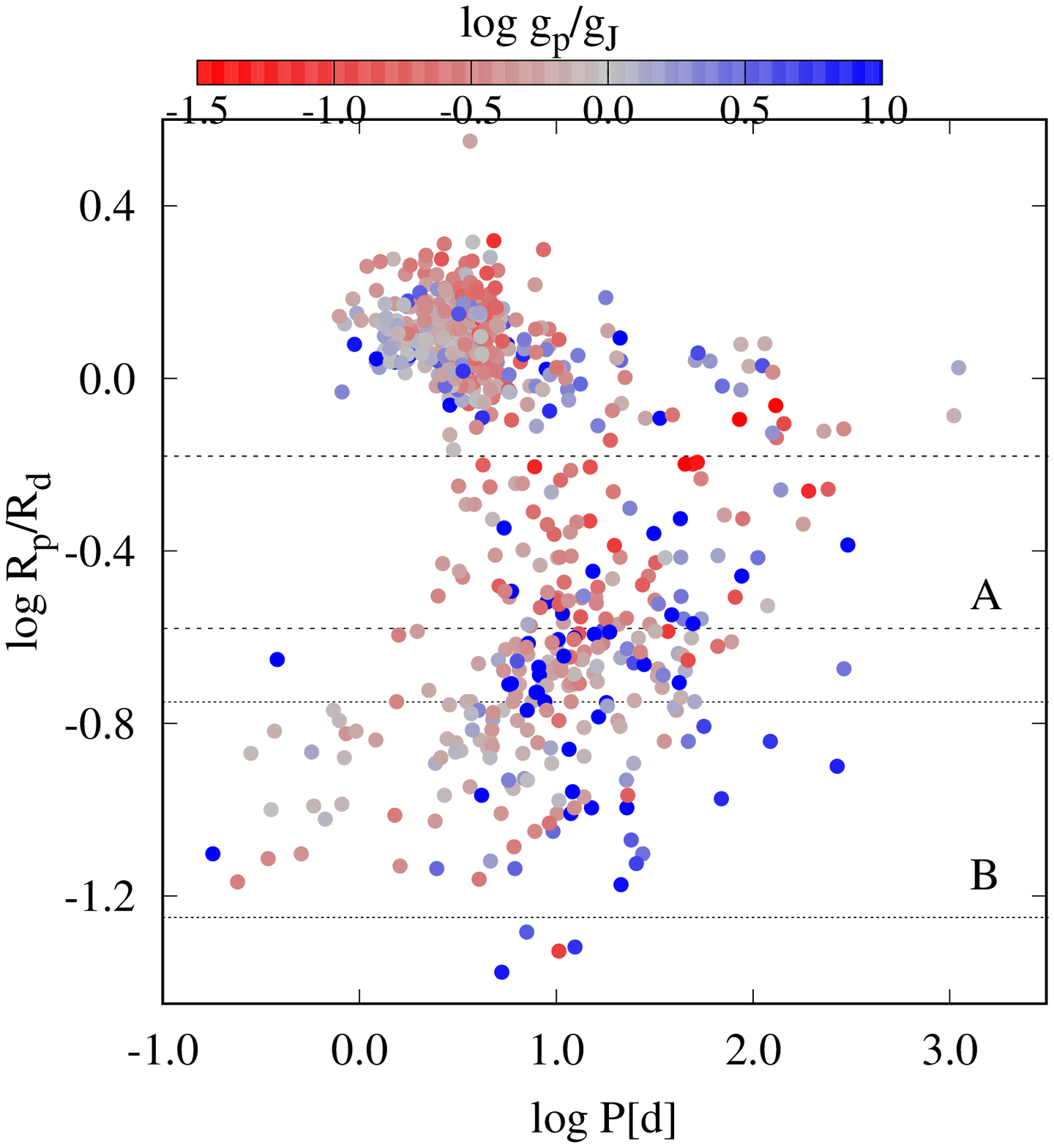}\hskip15mm%
\includegraphics[viewport=143 261 511 618,  width=4.3cm]{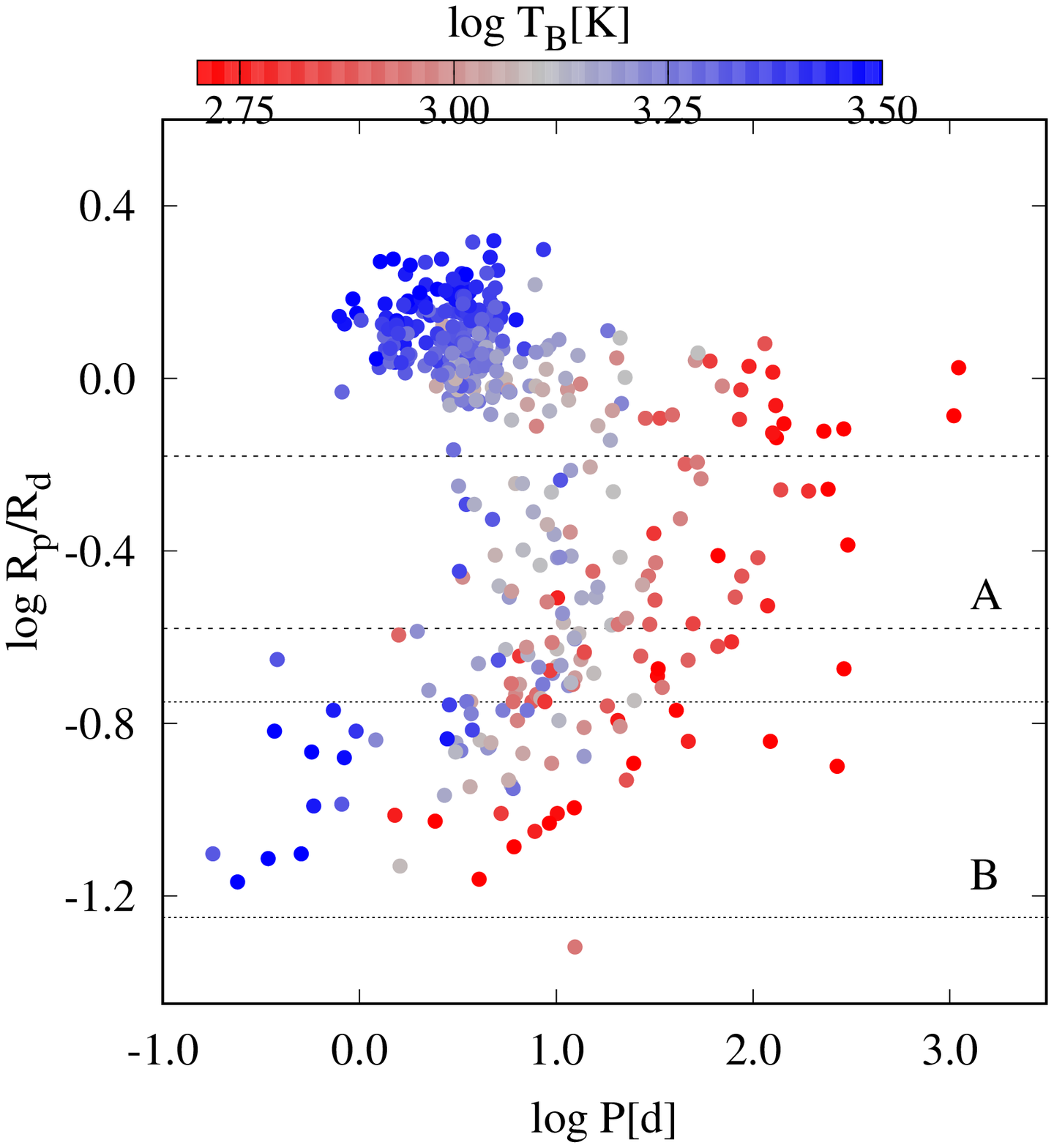}%
\vskip18mm%
\hskip6.0mm\includegraphics[viewport=107 215 396 323,width=4.3cm]{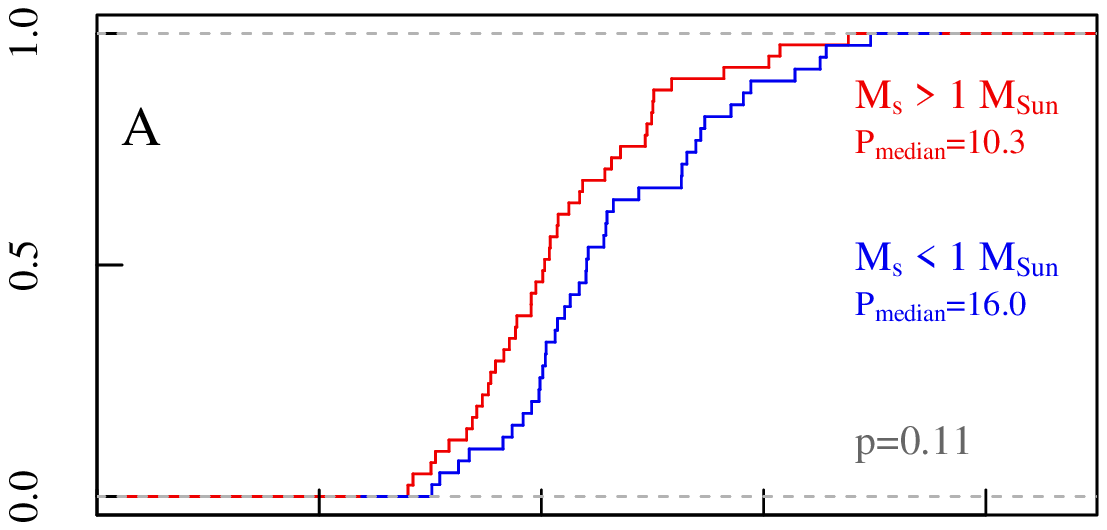}\hskip15mm%
\includegraphics[viewport=107 215 396 323,width=4.3cm]{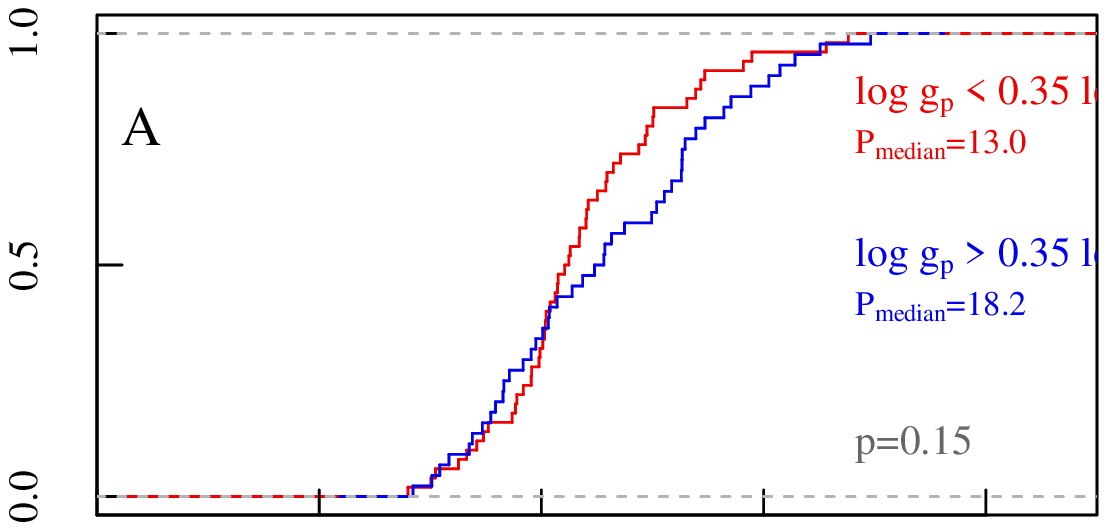}\hskip15mm%
\includegraphics[viewport=107 215 396 323,width=4.3cm]{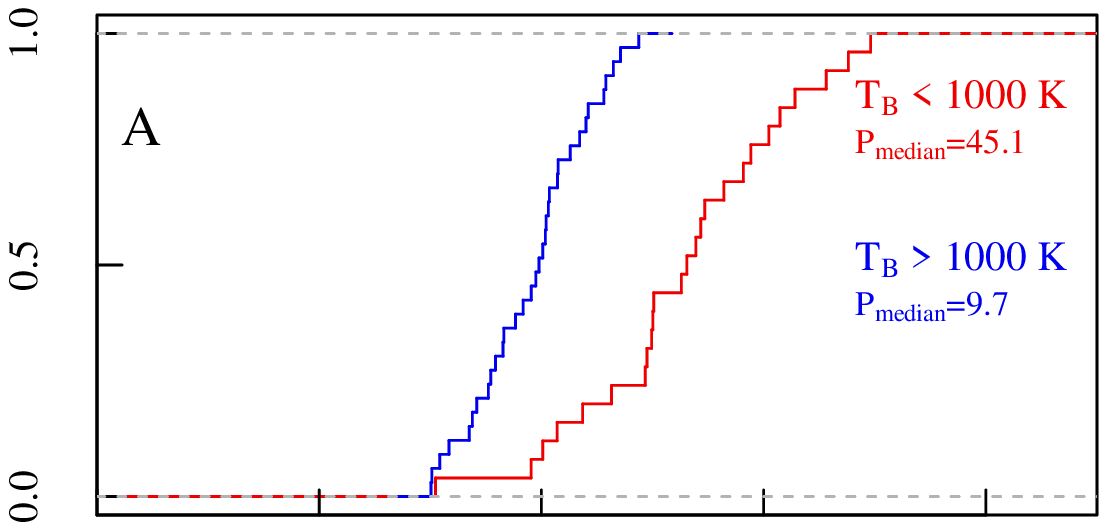}\hskip0mm%
\vskip8mm
\hskip6.0mm\includegraphics[viewport=107 181 396 323,width=4.3cm]{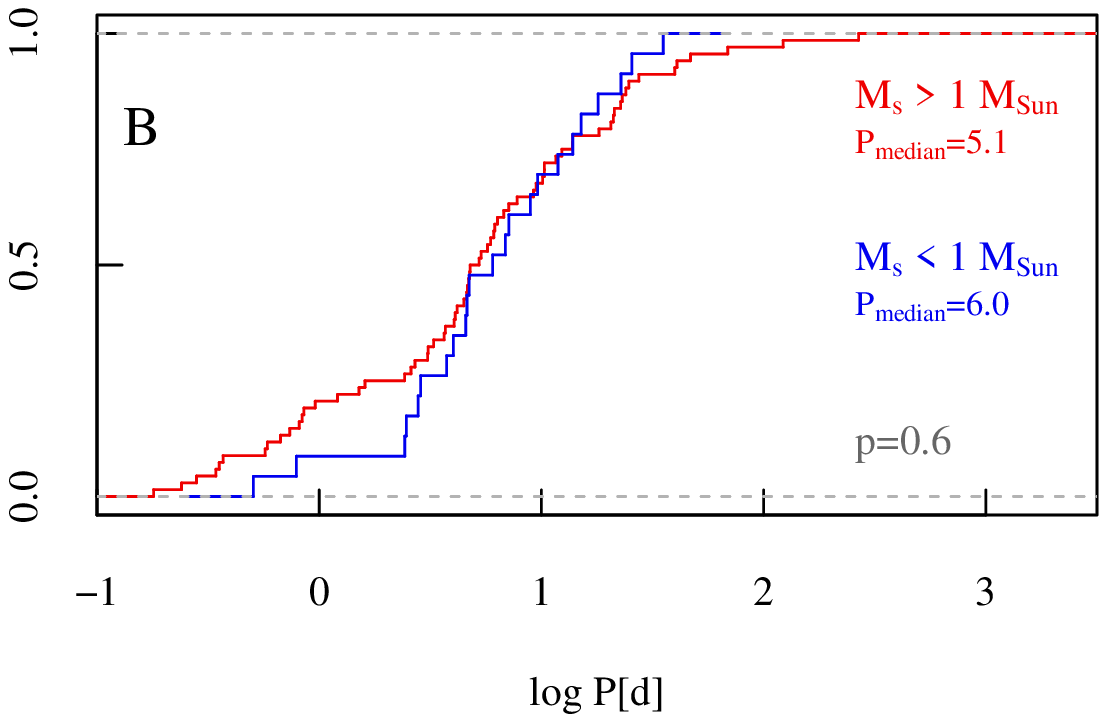}\hskip15mm%
\includegraphics[viewport=107 181 396 323,width=4.3cm]{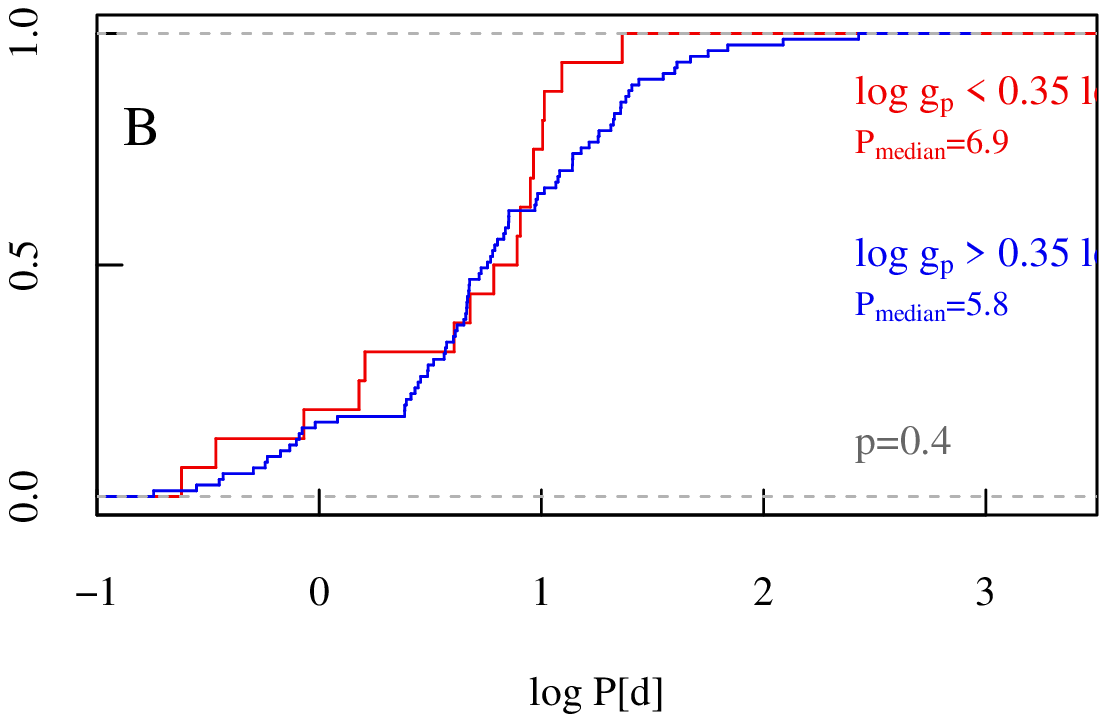}\hskip15mm%
\includegraphics[viewport=107 181 396 323,width=4.3cm]{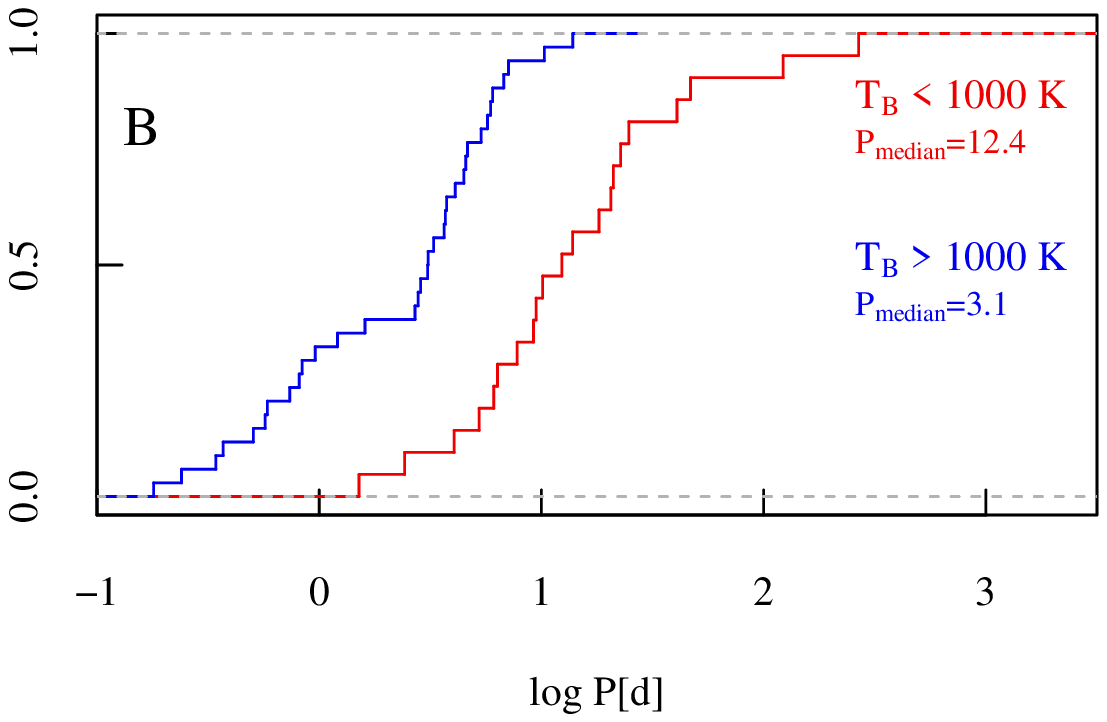}%
\vskip10mm
\caption{Selected examples for inconclusive parameter dependencies of the boundaries of the sub-Jupiter desert: stellar $M_*$ mass (left column), $\log g_p - \log g_J$ relative gravity at the top of the planetary atmosphere, $T_B$ black-body equivalent temperature of the planet. }
\end{figure*}


\subsection{The dependence on stellar temperature}

The boundary of the desert most significantly depends on \teff{} stellar temperature (Fig. 1 left column), the K-S test value is $p=0.0002$. {In all figure panels, we give the median values of the distribution, which is roughly 9 and 21 days for the subsets of colder and warmer host stars.} Besides, more than 60\%{} of planets around $T_{\rm eff}<5600$~K stars have shorter periods than 10 days, but only 10\%{} of planets around $T_{\rm eff}>5600$~K stars. The effect is located on the boundary of the desert only, there is no difference in the period distribution of the small planets around the warmer/colder host stars in the control region B ($p=0.6$, bottom panel).

Since shorter period planets are reasonably easier to discover, we see no other explanation for the lack of planets in the $<10$ day period range than the stellar effective temperature affects the boundary period of the sub-Jupiter desert.

\subsection{The metallicity dependence}

The second significant dependence of the period boundary of the sub-Jupiter desert is the stellar $[M/H]$ metallicity. We distinguished the high-metallicity and low-metallicity groups are separated at the median metallicity $[M/H]=0.05$ in the A region of exoplanets. 75\%{} of the planets in the high-metallicity group have orbital periods of less than 10-11 days, which is true only for less than 20\%{} of low-metallicity planets. Again, no correlation like this has been observed in the B region of small planets ($p=0.3$), proving the selectivity of metallicity in forming the desert (Fig. 1 middle column).

This correlation has been also found by Dong et al. (2017) and Petigura et al. (2018), and is compatible with the predictions accounting for metallicity-dependent planet photoevaporation (Owen \&{} Lai, 2018). Here we also prove that the period distribution of small planets does not follow a metallicity-dependence. Metallicity is selective only for the period boundary of the sub-Jupiter desert, not for smaller planets, and hence, it is a diagnostic of the process that forms the desert.

\subsection{The $\log g$ dependence}

There also exists a dependence of the period boundary on the stellar $\log g$, allowing planets around stars with higher $\log g$ to go deeper into the desert. The most significant difference is around the 10-day orbital period: 10\%{} versus 60\%{} of planets around low/high $\log g$ stars have shorter periods than 10 days (Fig. 1 right column).

However, the combined parameter dependence is somewhat degenerated: we saw that the planets which can go deeper into the desert (i.e. the minimum orbital period boundary is lower) should have a low temperature host star with higher values of $\log g$. But on the Main Sequence, stars with lower effective temperature have naturally higher $\log g$ (e.g Gazzano et al. 2013), so the parameter dependence on $T_{\rm eff}$ and $\log g$ can explain each other. It remains hidden in the data whether the $\log g$ dependence of the period boundary is a consequence of its temperature dependence, or is also connected to dynamical processes which enhance the naturally existing correlations.


\subsection{The mass of the host star}

\begin{figure}
\hskip0.8cm\includegraphics[viewport=108 230 400 325, width=7.3cm,clip=T]{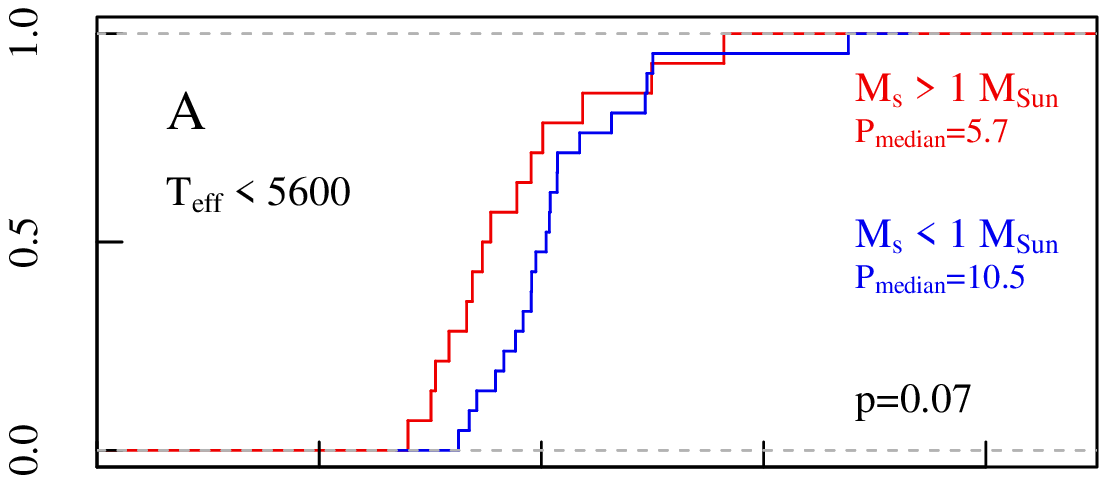}\\ \vskip8mm 
\hskip0.8cm\includegraphics[viewport=108 125 400 325, clip=T,  width=7.3cm]{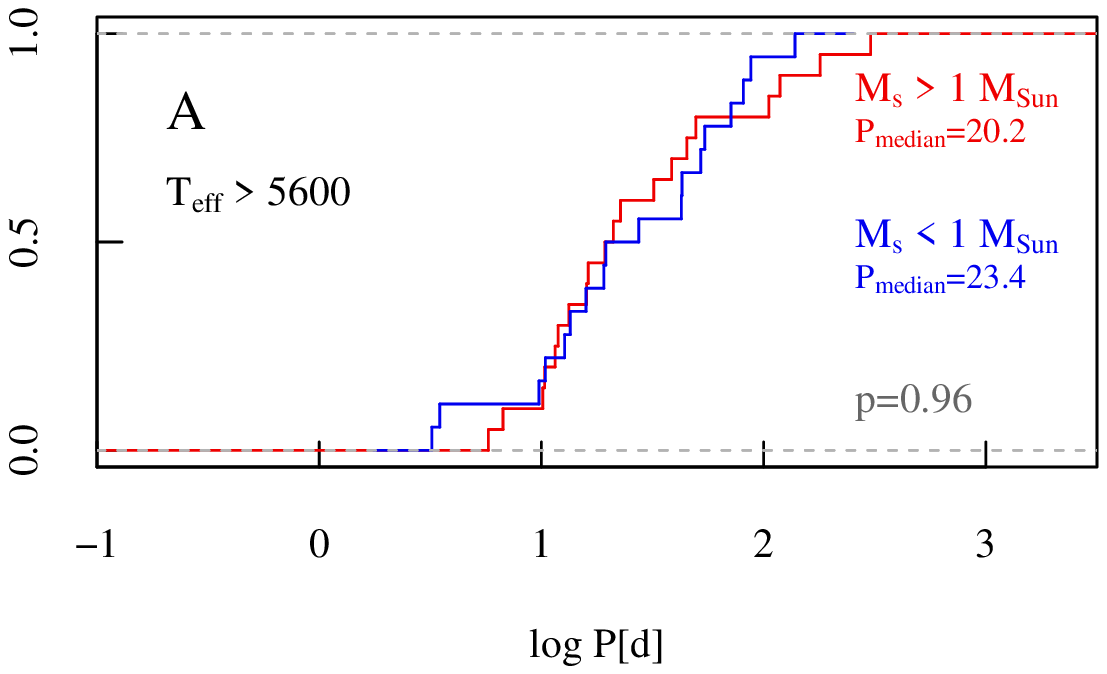}%
\caption{The dependency of the desert boundary on stellar mass in region A. The sample beyond Fig. 2 left column middle panel, is also split by effective temperature: the distributions around colder and hotter stars are shown in upper and lower panels, respectively.}

\hskip0.8cm\includegraphics[viewport=108 125 400 325, width=7.3cm,clip=T]{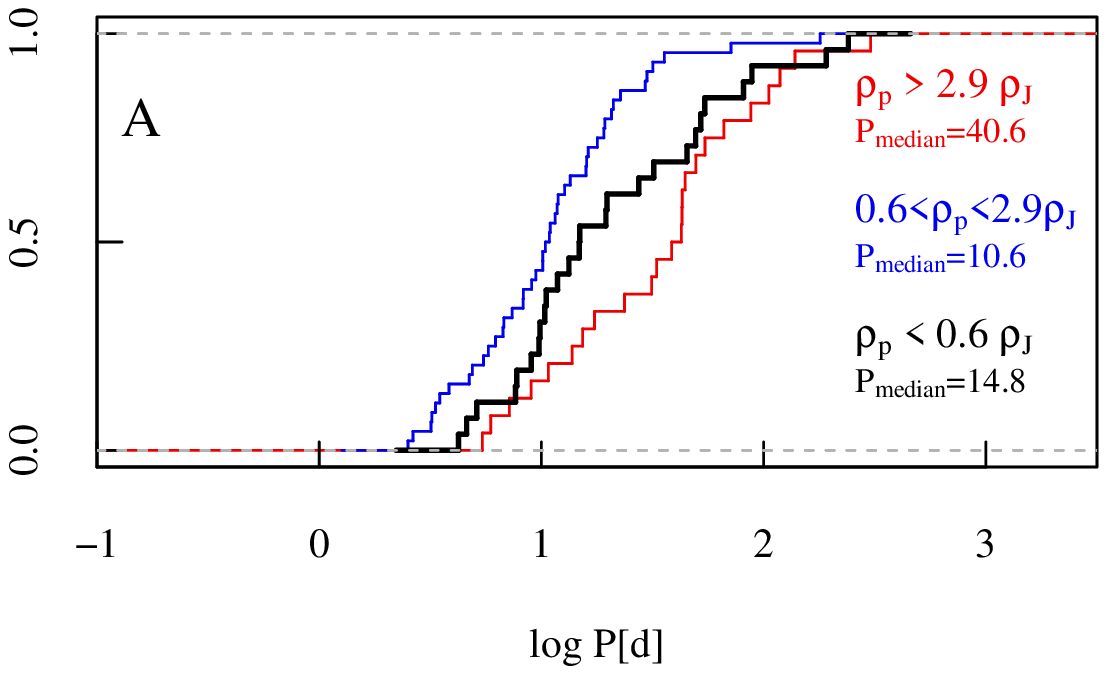}
\caption{Period distribution of three disjoint subsets of exoplanet, in the lower quartile (black), upper quartile (red) and interquartile (blue) range of planet density.}
\end{figure}

{
The dependence of the period boundary on stellar mass shows a $p$ value of $0.11$, {\modified bringing up the idea} that planets around more massive stars {\modified could} represent shorter minimum periods, and they {\modified could} go deeper into the desert. This finding {\modified would be} in a seeming contradiction with the {\modified result} about \teff{}, which gives a preference for host stars with lower temperature, which have a smaller stellar mass. To break this antagonism, the anonymous Referee suggested trying to separate the stellar mass effect from the temperature dependence.

Fig. 3 shows what is seen when we split the sample at the median $T_{\rm eff} \approx 5600$~K. In the case of colder stars, the stellar mass is recognized to be selective with an increased significance ($p=0.07$), but for hotter stars than 5600~K, the dependence completely vanishes. Thus, we conclude that the stellar mass has an influence to the desert boundary only in the case of host stars with $T_{\rm eff} < 5600$~K.
}

\subsection{The dependence on planet parameters}

We found no conclusive dependence of the period boundary on most planet parameters, and parameters that describe the interactions of the planet and the star on the current orbit. We involved the size of the Roche-lobe, and the actual tidal forces on the surface of the star (characterised by $M_s*r_p/a^3$), and the method that proved the effect of stellar parameters gave the negative results in all cases. Here we show the example of planet surface gravity (scaled by that of Jupiter, $\log g_p/g_J$) as an example in Fig. 2 middle column. The K-S test gave insignificant results here. What we see is a complex pattern, showing a group of $\log g_p/g_J>0$ planets at around the lower characteristic period of the desert, but well outside it, orbiting mostly with $>10$ day periods. Most part of the boundary of the desert, the A region, is characterised by an overpopulation of planets with $g_p < g_J$ (reddish points), which is also characteristic for inflated hot Jupiters (near the top of the hot Jupiter clump). This complex pattern should be a result of different processes during planet formation, including the runaway gas accretion above 13--15 Earth-mass, and photoevaporation of planets very close to the star. Indeed, the most inflated planets (red dots) are far from the boundaries of the desert, although they are the hottest planets; suggesting they all have a somewhat destroyed atmosphere.

In Fig. 2 right column, the map of the planet surface temperature is shown at the boundaries of the desert. Due to the Lambert-law, there is a natural strong dependence between the orbital period and the equilibrium temperature, which makes difficult the detection of possible selection effects above the natural correlations. {(Therefore, a $p$ value does not help here.)} However, comparing the histograms in regions A and B, we see that the distribution of colder ($T_B<1000$~K) planets is very similar near at the exoplanet boundary and below the desert: the median period is only a little bit more at the boundary, and the slope of the distributions are rather similar. However, the distribution of hot planets ($T_B<1000$~K) is strongly truncated by the desert, and the slope is very steep at the boundary. This suggests that the irradiation suffered by the planet is a dominant factor in shaping the sub-Jupiter desert.

{
We also plotted the distribution of average densities of exoplanets. The map (not shown here) is visually noisy and is roughly similar to the map of surface gravity. An interesting feature of this distribution is the lack of inflated planets at the boundary of the desert, which again may be a diagnostic for photoevaporation.

To get to the firm conclusion, we compare the period distributions of planets in subsets with different $\rho_{\rm p}$ planet density. In Fig. 4, the red, blue and black curve represent the planet clusters in the upper quartile, interquartile and lower quartile range of $\rho_{\rm p}$. We find moderately inflated planets orbiting closer to the star (the blue curve is closer than the red curve). The surprising feature is that on closest orbit to the star, we do not see any of the most inflated planets; even, the most inflated planets are distributed farther from the star (black curve) than the moderately inflated ones (blue curve). The K-S test gives a $p=0.12$ value for the differing distributions beyond the blue and black curves, and a Cramer-von Mises test (most sensitive to local differences, Feigelson and Babu, 2011) gives a very significant value of $p=0.02$.  
}

\section{Conclusions}


{
Here we have shown that the boundary of the sub-Jupiter desert depends on the fundamental stellar parameters, including the temperature, metallicity, $\log g$ and probably the stellar mass, in the order of decreasing significance. We suggest that several dependencies of the desert ($M/H$ metallicity, \teff{}, $\log g$, $M_*$ stellar mass) are related to the predominance of photoevaporation processes. The explanation of how the desert boundary depends on \teff{} (as a source of incident flux and also, incident UV irradiation) is straightforward in this context. 

Also, the metallicity effect can be explained by photoevaporation. It is known that stars with higher metallicity tend to host more massive debris disks (G\'asp\'ar et al. 2016). It is plausible to assume that the total mass of planetesimals were more in high-metallicity disks, and the core of planets in high-metallicity systems is also more massive. Planets with more massive cores are more resistant against photoevaporation, they can found to go deeper in the desert (Owen \&{} Lai 2018).

The observed pattern of planet densities is also explained if the photoevaporation is the dominant factor in forming the desert. We see a lack of the most inflated planets on the closest orbits, probably because the most inflated planets on very close orbits have been evaporated by now. If planets under a critical density loose a part of the atmosphere, the density will increase which can stabilize the planet again, and may be this is what we see in Fig. 4.}

According to this conclusion, the non-dependence of the period boundary on the tidal forces and the surface $\log g_p$ of the planet gives an observational support for a lower effect of tidal or dynamical interactions in forming the sub-Jupiter desert. The odd distributions of planets with low and high surface temperature also support the increased role of photoevaporation to tidal interactions.


\section*{Acknowledgements}
This research was founded by the
Hungarian NKFI Grants GINOP-2.3.2-15-2016-00003 and K-119517 of the Hungarian
National Research, Development and Innovation Office, and
by the City of Szombathely under agreement No. S-11-1027.
Authors thank to K. Nagy and A. Murvai for their suggestions and discussions.

\end{document}